\documentclass{elsart}
\usepackage{graphicx,natbib,amsmath}
\journal{International Review of Financial Analysis}
\newcommand{\avg}[1]{\langle #1\rangle}
\newcommand{\vek}[1]{\boldsymbol{#1}}
\newcommand{\req}[1]{(\ref{#1})}
\newcommand{\mef}{m_{\text{ef}}}
\newcommand{\id}{\mathrm{id}}
\newcommand{\C}{\mathsf{C}}
\begin{document}
\begin{frontmatter}
\title{How to quantify the influence of correlations
on investment diversification}
\author[fr]{Mat\'u\v s Medo\corauthref{cor}},
\corauth[cor]{Corresponding author. Tel. +41 26 300 9139.}
\ead{matus.medo@unifr.ch}
\author[ho,fr,che]{Chi Ho Yeung},
\ead{phbill@ust.hk}
\author[fr,che]{Yi-Cheng Zhang}
\ead{yi-cheng.zhang@unifr.ch}

\address[fr]{Physics Department, University of Fribourg,
P\'erolles, 1700~Fribourg, Switzerland}
\address[ho]{Department of Physics, The Hong Kong University of
Science and Technology, Hong Kong, China}
\address[che]{Lab of Information Economy and Internet Research,
University of Electronic Science and Technology of China,
610054~Chengdu, China}

\begin{abstract}
When assets are correlated, benefits of investment
diversification are reduced. To measure the influence of
correlations on investment performance, a new quantity---the
effective portfolio size---is proposed and investigated in
both artificial and real situations. We show that in most cases,
the effective portfolio size is much smaller than the actual
number of assets in the portfolio and that it lowers even
further during financial crises.
\end{abstract}

\begin{keyword}
Mean--Variance portfolio \sep Kelly portfolio \sep
diversification \sep correlations
\end{keyword}

\end{frontmatter}

\section{Introduction}
Investment optimization, pioneered by Markowitz~\citep{Ma52},
is one of the key topics in finance. It aims at simultaneous
maximization of the investor's capital and minimization of the
risk of unfavorable events. When these goals are mathematically
formalized, they give rise to various methods of portfolio
optimization. In this paper we focus on the Mean--Variance
method~\citep{Ma52} and the maximization of the expected
exponential growth rate~\citep{Ke56}. For a thorough review of
modern portfolio theory see~\citep{EGBG06}.

Most portfolio optimization strategies result in investment
diversification as it allows investors to decrease their
exposure to the risk of single assets~\citep{Lin65}. In
consequence, diversification represents a major issue in modern
finance theory and practice~\citep{Ma91}. The basic premise is
that with sufficient diversification one is able to reduce
fluctuations to an acceptable level. However, when assets are
correlated, the improvement of investment performance achieved
by diversification is reduced~\citep{EG77}. Since asset
correlations are ubiquitous, ranging from correlations between
stocks in one stock market~\citep{St87} to correlations between
different investment types in
different countries~\citep{Jo85,Meric08,Oli08}, it is important
to investigate their influence on diversified
portfolios~\citep{HeRo94,Po05}. Worse, during financial crises
like the one we currently endure, we often observe that when we
need diversification the most it often fails us, leading to
massive losses~\citep{Scholes00}.

In this paper we attempt to quantify how correlations reduce the
benefits of diversification. To achieve this we propose a new
measure, the effective size of a diversified portfolio, which
is based on a comparison with a fictitious portfolio of
uncorrelated assets. We apply this idea for two different
optimization strategies (the Mean--Variance portfolio and the
Kelly portfolio) and obtain analytical expressions for their
effective sizes. We show that this number is often small, even
for a portfolio constructed from a very large number of stocks.
The achieved results are also used to study real market data
(daily prices of stocks from the Dow Jones Industrial Average
and the S\&P~500), showing that the effective portfolio size
varies significantly for different sets of stocks and different
times.

\section{Correlations and the Mean--Variance portfolio}
First we introduce the notation used in this paper. If the
initial and final asset values are $w_0$ and $w_1$ respectively,
the asset return during the given time period is defined as
$R:=(w_1-w_0)/w_0$. If we follow the value of asset $i$ over
many subsequent time periods, it is possible to define the
average return $\mu_i:=\avg{R_i}$, the return variance
$V_i:=\avg{R_i^2}-\avg{R_i}^2$ and the standard deviation
$\sigma_i:=\sqrt{V_i}$ ($\avg{x}$ denotes the average value of
$x$). To measure correlations of returns, the Pearson's formula
is the standard tool. For assets $i$ and $j$ it reads
\begin{equation}
\label{Pearson}
C_{ij}:=\frac{\avg{R_iR_j}-\avg{R_i}\avg{R_j}}{\sigma_i\sigma_j},
\end{equation}
by definition $C_{ij}\in[-1;1]$ and $C_{ii}=1$. When returns
$R_i$ and $R_j$ are independent,
$\avg{R_iR_j}=\avg{R_i}\avg{R_j}$ and hence $C_{ij}=0$. The same
holds when one of the assets is risk-free, \emph{i.e.} its
return has zero variance. (Although in practice each investment
carries a certain amount of risk, short term government-issued
securities are often used as proxies for risk-free assets.) We
assume that a risk free asset is available and has zero return.
Correlation values of all assets form the correlation matrix
$\C$.

To construct a portfolio, the investor has to divide the current
wealth $W$ among $M$ available assets. This division can be
characterized by investment fractions: $f_i$ is the fraction of
wealth invested in asset $i$ ($i=1,\dots,M$). Assuming unit
initial wealth, after one time period the investor has wealth
\begin{equation}
\label{W_1}
W_1=1+\sum_{i=1}^M f_iR_i.
\end{equation}
Here $R_i$ is the return of asset $i$ during the period. When
investment fractions are fixed, the investor's wealth follows a
multiplicative stochastic process and after $T$ periods it
becomes
\begin{equation}
\label{W_T}
W_T=\prod_{t=1}^T \Big(1+\sum_{i=1}^M f_iR_{i,t}\Big).
\end{equation}
Here $R_{i,t}$ is the return of asset $i$ in the period $t$.

The Mean--Variance approach to portfolio optimization has been
proposed by Markowitz~\citep{Ma52}, for later discussions see
\citep{Ma72,Ma91,EGBG06}. Despite its flaws (\emph{e.g.}, only
the expected return and its variance are used to characterize a
portfolio) it is still a~benchmark for other optimization
methods. For any portfolio, we can compute the expected return
$R_P:=\avg{W_1}-1$ and the variance
$V_P:=\avg{W_1^2}-\avg{W_1}^2$ as
\begin{equation}
\label{RandV}
R_P=\sum_{i=1}^M f_i\mu_i,\quad
V_P=\sum_{i,j=1}^M f_if_jC_{ij}\sigma_i\sigma_j.
\end{equation}
The optimal portfolio is defined as the one that minimizes $V_P$
for a~given $R_P$ (equivalently, one can maximize $R_P$ with
$V_P$ fixed). To focus purely on the influence of correlations
on the optimization process, we assume that all $M$ games are
identical: $\mu_i=\mu$ and $\sigma_i=\sigma$. The minimal
portfolio variance $V_P^*$ then has the form
\begin{equation}
\label{MV-V_P}
V_P^*(R_P,M,\C)=
\frac{\sigma^2 R_P^2}{\mu^2\sum_{i,j=1}^M(\C^{-1})_{ij}}
\end{equation}
where $\C^{-1}$ is the inverse of the correlation matrix $\C$.
The larger is the expected portfolio return, the larger is the
optimal variance.

Now we look at Eq.~\req{MV-V_P} from a new perspective: we
compare it with the optimal variance of the portfolio of $m$
uncorrelated assets with identical mean returns and variances
which is $V_P^*(R_P,m,\id_m)$ where $\id_m$ is $m\times m$
identity matrix. We can introduce the effective size of the
correlated portfolio, $\mef$, by comparing the two optimal
variances. The equation
$V_P^*(R_p,M,\C)=V_P(R_p,\mef,\id_{\mef})$ can be solved with
respect to $\mef$, yielding
\begin{equation}
\label{m_ef-MV-general}
\mef=\sum_{i,j=1}^M(\C^{-1})_{ij}.
\end{equation}
In other words: when investing in $M$ correlated assets, the
portfolio variance is the same as for $\mef$ uncorrelated
assets. Note that $\mef$ depends only on the correlation matrix
and not on the portfolio parameters $\mu,\sigma,R_P$. A similar
sum over the elements of the inverse correlation matrix arises
in the expression for the magnetic susceptibility in the Ising
model (which is a prominent model of magnetism in theoretical
physics).

To better understand the new concept, we examine the special
case of uniform correlations between the assets: $C_{ij}=C$ for
$i\neq j$, $C\in[0;1]$. Then $\C^{-1}$ has the form
\begin{equation}
\label{inverseC}
(\C^{-1})_{ii}=\frac{1+(M-2)C}{(1-C)[1+(M-1)C]},\quad
(\C^{-1})_{ij}=\frac{-C}{(1-C)[1+(M-1)C]}.
\end{equation}
and Eq.~\req{m_ef-MV-general} consequently simplifies to
\begin{equation}
\label{m_ef-MV-C}
\mef=\frac{M}{1+(M-1)C}.
\end{equation}
Now we can get some intuition about the effective
portfolio size: when $C=1$ (perfectly correlated assets),
$\mef=1$; when $C=0$ (uncorrelated assets), $\mef=M$; when
$C=-1$ and $M=2$, the portfolio variance can be totally
eliminated and $\mef\to\infty$. A remarkable consequence of
Eq.~\req{m_ef-MV-C} is that in the limit $M\to\infty$ we obtain
$\mef=1/C$. This means that diversification into arbitrary many
assets with mutual correlation $C$ is equivalent to investment
in only $1/C$ uncorrelated assets.

Another interesting case is a~block diagonal matrix $\C$ which
has square matrices $\C_1,\dots,\C_N$ along the main diagonal
and the off-diagonal blocks are zero matrices (such a form of
$\C$ is an extreme case of the sector structure discussed in
Sec.~\ref{sec-sectors}). It can be shown that
Eq.~\req{m_ef-MV-general} then yields
\begin{equation}
\label{m_ef-blocks}
\mef=\mef(\C_1)+\dots+\mef(\C_N).
\end{equation}
Thus, the effective portfolio size is the sum of effective sizes
for each block $\C_i$ separately.

It is instructive to compare the results obtained above with the
simple investment distributed evenly among all assets. If we
assume identical returns and variances of the assets,
Eq.~\req{RandV} simplifies to $R_P=Mf\mu$,
$V_P=\sigma^2f^2\sum_{i,j=1}^M C_{ij}$, where $f$ is the
fraction of wealth invested in each single asset. The desired
value of $R_P$ now determines both $f$ and $V_P$. By comparison
with the variance of a portfolio of uncorrelated assets we
obtain the effective size of the even investment in the form
\begin{equation}
\label{m_ef-even}
\mef'=\frac{M}{1+(M-1)\avg{C}}
\end{equation}
where $\avg{C}$ is the average of the off-diagonal elements of
$\C$, the prime symbol indicates that the even investment is
considered. We shall see (Sec.~\ref{sec-real}) that $\mef'$ is
often significantly smaller than $\mef$.

We emphasize that the effective portfolio size $\mef$ is
different from the inverse participation ratio (also called the
Herfindahl index) which is defined as
$\mathrm{IPR}=1/\sum_{i=1}^M f_i^2$ ($f_i$ is the fraction of
wealth invested in asset $i$). While the former quantifies the
influence of correlations, the latter quantifies how unevenly
wealth is distributed among the assets.

\section{Correlations and the Kelly portfolio}
The Kelly portfolio~\citep{Ke56} maximizes the investment
performance in the long run and has several interesting
mathematical properties~\citep{Br00,FiWh81,Th00}. Its
applicability was investigated in many different
situations~\citep{Ma76,our}, including those with limited
information~\citep{Br96,our2,Smimou08}, and it has been
successfully used in real financial markets~\citep{Th00}. In the
following paragraphs we show how the effective portfolio size
can be introduce for the Kelly portfolio.

Since $W_T$ given by Eq.~\req{W_T} follows a multiplicative
random walk, investor's wealth grows exponentially as
$W_T=\exp[GT]$. Consequently, the long-run exponential growth
rate $G$ can be written in the form
\begin{equation}
\label{G-def}
G:=\lim_{T\to\infty}\frac1T\ln\frac{W_T}{W_0}.
\end{equation}
Maximization of this quantity is the main criterion for the
construction of the Kelly portfolio. Using the law of large
numbers, it can be shown that $G=\avg{\ln W_1}$ where $W_1$ is
the wealth after one step. Assuming investment into $M$
simultaneous games we have
\begin{equation}
\label{G}
G=\bigg\langle\ln\Big[1+\sum_{i=1}^M f_iR_i\Big]\bigg\rangle.
\end{equation}
The simplest case is one risky game with binary outcomes: $R=+1$
with the probability $p$ and $R=-1$ with the probability $1-p$.
The maximization of $G$ then yields the optimal fraction
\begin{equation}
\label{f-Kelly}
f^*=2p-1
\end{equation}
which is the celebrated Kelly criterion (if short positions are
not allowed, $f^*=0$ for $p<1/2$: it is optimal to abstain from
the game). Eq.~\req{G}, allowing no analytical solution for
$M\geq5$, can be maximized by numerical techniques~\citep{Wh07}
or by analytical approximations~\citep{our2} as we do below.

\subsection{The effective size of the Kelly portfolio}
To investigate the effect of asset correlations on the Kelly
portfolio, we consider $M$ individual assets with the
correlation between assets $i$ and $j$ computed by
Eq.~\req{Pearson} and labeled as $C_{ij}$. Differentiation of
Eq.~\req{G} with respect to $f_i$ yields
\begin{equation}
\label{hetero-G}
\sum_{\vek{R}}\frac{P(\vek{R}) R_i}{1+\sum_{j=1}^M f_j R_j}=0
\end{equation}
where $P(\vek{R})$ is the probability of a given vector of
returns $\vek{R}=(R_1,\dots,R_M)$ and the summation is over all
possible $\vek{R}$ (when returns are continuous, integration
must be used instead). For $M>4$, this equation has no
analytical solution. Assuming that the investment return
$\sum_{i=1}^M f_i R_i$ is small (which is plausible if the
considered time period is short), we can use the expansion
$1/(1+x)\approx 1-x$ to obtain
\begin{equation}
\label{1st_approx-Kelly}
\sum_{j=1}^M f_j\avg{R_iR_j}=\avg{R_i}\qquad
(i=1,\dots,M).
\end{equation}
This set of linear equations gives the first approximation to
$f_i$: when $\avg{R_i}$ and $\avg{R_iR_j}$ are known, the
optimal investment fractions $f_i^*$ follow. A higher order
expansion of $1/(1+x)$ results in higher order cross terms of
returns which are generally difficult to compute.

As before, to focus on the influence of correlations we assume
identical return distributions of the assets: we set
$\avg{R_i}=\mu$ and $\sigma_i=\sigma$ for $i=1,\dots,M$,
consequently $\avg{R_iR_j}=\mu^2+\sigma^2 C_{ij}$. After
substitution to Eq.~\req{hetero-G}, the optimal investment
fractions are
\begin{equation}
\label{kelly-fvector}
\vek{f}^*=\frac{\mu\,(\C^{-1}\vek{1})}
{\sigma^2+\mu^2\sum_{i,j=1}^M(\C^{-1})_{ij}}
\end{equation}
where $\vek{1}$ is the $M$-dimensional vector with all elements
equal to $1$. The effective portfolio size $\mef$ is again
obtained by a comparison with a portfolio of uncorrelated
assets. One way to do this is by comparing the total invested
wealth in both cases. That is, $\mef$ is defined as the solution
of
\begin{equation}
\label{m_ef-def}
\sum_{i=1}^M f_i^*(M,\C)=\mef\, f^*(\mef,\id_{\mef}).
\end{equation}
Using Eq.~\req{kelly-fvector} we readily obtain
\begin{equation}
\label{m_ef-Kelly-general}
\mef=\sum_{i,j=1}^M(\C^{-1})_{ij}
\end{equation}
which is exactly the same expression for $\mef$ as
Eq.~\req{m_ef-MV-general} in the Mean--Variance approach. We can
conclude that the effective portfolio size is a common quantity
for these two optimization schemes. Nevertheless, this exact
correspondence is valid only when the first order approximation
is used to solve Eq.~\req{hetero-G}.

It is also possible to define the effective number of assets
$\mef$ differently: as the number of uncorrelated assets when
the expected exponential growth rate $G(\mef,\id_{\mef})$ is
equal to the growth rate $G(M,\C)$ with $M$ correlated assets.
This definition is similar to the one introduced above and it
also yields similar results. For practical reasons (the total
investment fraction is easier to handle analytically than the
exponential growth rate) we confine our analysis to the former
definition.

\subsection{Special case with identical correlations}
\begin{figure}
\centering
\includegraphics[scale=0.28]{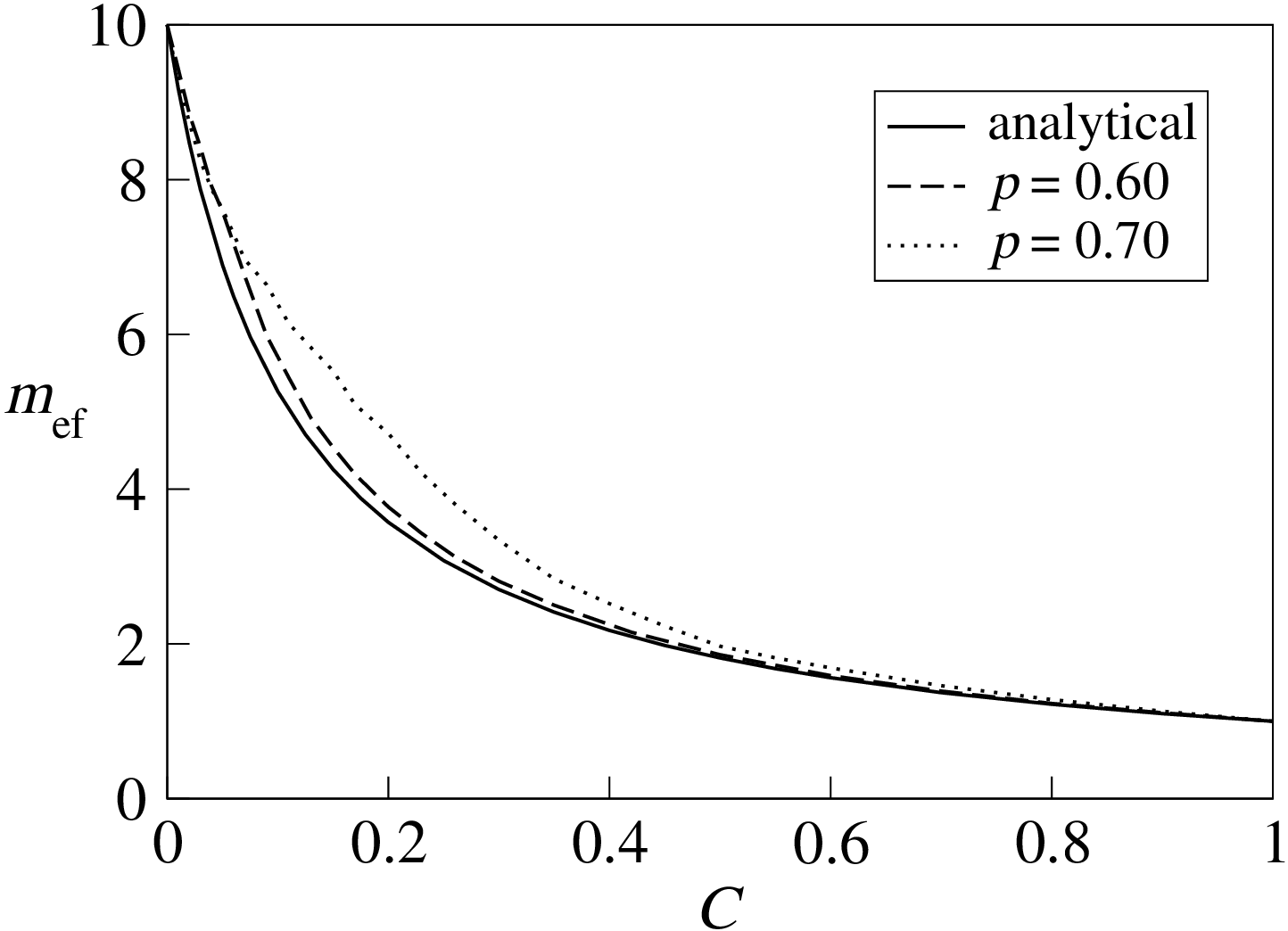}
\caption{The effective size of the Kelly portfolio: a comparison
of the approximate result Eq.~\req{m_ef-Kelly-C} with numerical
treatment of Eq.~\req{homo-G} for various winning probabilities.
The total number of assets is $M=10$.}
\label{fig:m_ef}
\end{figure}

Assuming identical asset correlations, $C_{ij}=C$ for $i\neq j$
($C\in[0;1]$), Eq.~\req{m_ef-Kelly-general} simplifies to
\begin{equation}
\label{m_ef-Kelly-C}
\mef=\frac{M}{1+(M-1)C}.
\end{equation}
We remind that while Eq.~\req{m_ef-MV-C} is exact, this result
is based on the first order approximation in Eq.~\req{hetero-G}.
To review its accuracy, we study the problem numerically. Since
with identical correlations all assets are equal, the optimal
investment is distributed evenly among them. Thus, maximization
of the exponential growth rate $G$ simplifies to a one-variable
problem and Eq.~\req{hetero-G} is replaced by
\begin{equation}
\label{homo-G}
\sum_{\vek{R}}\frac{P(\vek{R})\sum_{j=1}^M R_j}
{1+f\sum_{j=1}^M R_j}=0.
\end{equation}
To proceed, one needs to specify the joint distribution of
returns, $P(\vek{R})$. To do so we use simple assets with binary
outcomes: $R_i=1$ (which we label as $+_i$) and $R_i=-1$ (which
we label as $-_i$). To induce the correlations we use an
artificial hidden asset $h$ with the outcome $+_h$ with the
probability $p$ and $-_h$ with the probability $1-p$. Finally,
asset returns are drawn conditionally on the hidden asset
according to
\begin{equation}
\label{rules}
P(+_i\vert +_h)=p+(1-p)\sqrt{C},\quad
P(-_i\vert -_h)=1-p+p\sqrt{C},\quad
C\in[0;1].
\end{equation}
It can be shown that $P(+_i)=p$, $P(-_i)=1-p$, $C_{ij}=C$; thus
the proposed construction satisfies our requirement of identical
correlations. The distribution $P(\vek{R})$ is
\begin{equation}
\label{P(R)}
P(\vek{R})=P(+_h)\prod_{i=1}^M P(R_i\vert +_h)+
P(-_h)\prod_{i=1}^M P(R_i\vert -_h)
\end{equation}
and Eq.~\req{homo-G} can be solved, yielding the optimal
fraction $f^*$. Consequently, the definition relation
Eq.~\req{m_ef-def} allows us to interpolate $\mef$ for any
$M,p,C$ (interpolation is needed because the right side of
the definition equation can be numerically computed only for
integer $\mef$). In Fig.~\ref{fig:m_ef} we compare this result
with Eq.~\req{m_ef-Kelly-C}. As can be seen, when the investment
return is small ($p=0.60$ and less), the approximate result
performs well. When $p=0.70$, differences appear for
$C\in(0.1,0.4)$. These discrepancies are not surprising because
in that region, more than 80\% of wealth is invested, violating
the approximation used to solve Eq.~\req{hetero-G}.

\begin{figure}
\centering
\includegraphics[scale=0.28]{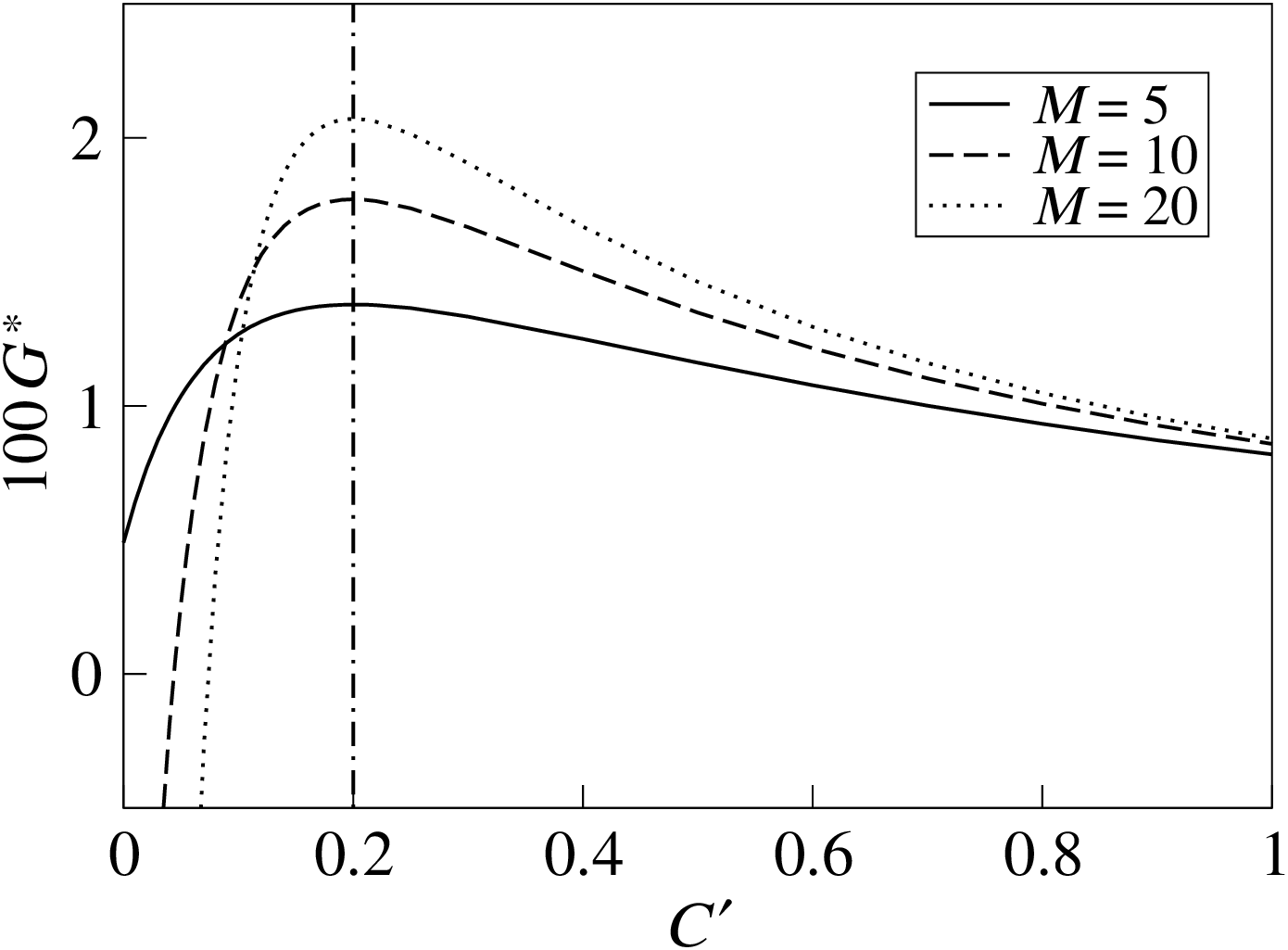}
\caption{The optimal exponential growth rate $G^*$ for an
investor assuming a wrong magnitude $C'$ of asset correlation.
The actual value $C=0.2$ is marked with the vertical dash-dot
line, the winning probability is $p=0.55$ (\emph{i.e.},
$\mu=0.1$).}
\label{fig:G}
\end{figure}

Finally, we use the established framework to investigate the
effect of wrong correlation estimation on portfolio performance.
We assume that all assets have pairwise correlation equal $C$
but the investor optimizes the investment assuming a different
value $C'$. In Fig.~\ref{fig:G}, the resulting growth rate $G^*$
is shown as a function of $C'$. As can be seen, underestimation
of correlations ($C'<C$) decreases investment performance
dramatically---a naive investor supposing zero correlations can
even end up with diminishing wealth. By contrast, a~similar
overestimation of $C$ results in only a mild decrease of the
growth rate.

\subsection{Estimates of the effective portfolio size}
\label{sec-sectors}
One can ask whether Eq.~\req{m_ef-Kelly-general} can be
approximated by a~simpler formula. Motivated by
Eq.~\req{m_ef-Kelly-C}, the natural guess is
\begin{equation}
\label{m_ef-avgC}
\mef=\frac{M}{1+(M-1)\avg{C}}.
\end{equation}
That means, we approximate diverse correlations by their
average. Since the resulting $\mef$ is the same as
Eq.~\req{m_ef-even}: this approximation is equivalent to
distributing the investment among the assets evenly.

In real markets, assets can be divided into sectors with
correlations higher between assets in the same sector than
between assets in different sectors. This sector structure can
be used to obtain an improved estimate of $\mef$. Assuming that
the assets can be divided into $N$ sectors, we denote the
intra-sector correlation between the assets from sector $I$ by
$\tilde C_{II}$ and the inter-sector correlation between the
assets from sectors $I$ and $J$ by $\tilde C_{IJ}$ (for indices
labeling sectors we use capital letters). Here $\tilde C_{II}$
and $\tilde C_{IJ}$ are simple averages ($I,J=1,\dots,N$)
\begin{equation}
\label{sectorCorr}
\tilde C_{II}=\frac1{M_I^2}\sum_{i, j\in I}C_{ij},\quad
\tilde C_{IJ}=\frac1{M_I M_J}\sum_{i\in I, j\in J}C_{ij},
\end{equation}
where $M_I$ is the number of assets in sector $I$. As a result,
an $N$-dimensional matrix $\tilde\C$ is formed. In
$\tilde C_{II}$ we sum also over diagonal elements of the asset
correlation matrix $\C$ as it is convenient for our further
computation.

Due to the simplifying assumption of identical intra-sector
correlations, the optimal investment fractions are identical
within a sector and the optimization problem simplifies to
$N$~variables $f_I$. Similarly to Eq.~\req{G}, the exponential
growth rate is
\begin{equation}
G=\bigg\langle\ln\Big[1+\sum_{I=1}^N
f_I\sum_{i\in I}R_i\Big]\bigg\rangle.
\end{equation}
By the same techniques as before, we obtain the estimate of
the effective portfolio size
\begin{equation}
\label{m_ef-sectors}
\mef=\sum_{I,J=1}^M(\tilde\C^{-1})_{I,J}.
\end{equation}
Its accuracy will be examined in the following section. For the
Mean--Variance approach, the sector-based estimate of $\mef$ is
the same.

\section{Correlations in real financial data}
\label{sec-real}
Here we test our results on real financial data, keeping two
goals in mind. First, we aim to investigate actual values of
the effective portfolio size. Second, we aim to examine the
accuracy of $\mef$ estimates derived above. We use prices of
stocks from the Dow Jones Industrial Average (DJIA) and the
Standard~\&~Poor's~500 (S\&P~500) which are well-known and
common indices consisting of 30 and 500 U.S. companies
respectively.

\subsection{Comparing index and stock variances}
\label{sec-avgC}
Using the daily data from the period 8th April 2004--14th
December 2007, we compute daily returns of the DJIA and the
included stocks by the formula $R(t+1)=(w(t+1)-w(t))/w(t)$, with
$w(t)$ denoting the adjusted closing index value on the trading
day $t$. The DJIA value is the sum of the prices of all
components, divided by the Dow divisor which changes with time.
This effect can be ignored because changes of the divisor are
mostly negligible.

During the given period, the variance of the DJIA daily returns
was $\sigma_{\mathrm{DJIA}}^2\approx 5.22\cdot 10^{-5}$, the
average variance of the daily returns of the DJIA stocks was
$\sigma_{\mathrm{S}}^2\approx 1.73\cdot 10^{-4}$. These two
figures contradict the assumption of zero correlations because
then the variance should scale with the number of assets $M$ as
$1/M$ (this follows also from Eq.~\req{MV-V_P} when $\C=\id_M$
is substituted). By dividing
$\sigma_{\mathrm{S}}^2/\sigma_{\mathrm{DJIA}}^2$ we obtain an
alternative estimate of the effective number of assets (this
estimate was used already in~\citep{GoKu04}). In our case it is
equal to $3.31$ which is much less than the total number of
stocks in the DJIA.

We can also estimate the effective number of assets using the
results of our analysis above. From the daily stock returns,
return correlations can be computed by Eq.~\req{Pearson},
resulting in $\avg{C}=0.322$. Together with the number of stocks
$M=30$, Eq.~\req{m_ef-avgC} yields $\mef\approx 2.90$. This is
in good agreement with the value $3.31$ obtained by a different
reasoning in the previous paragraph.

\subsection{The effective portfolio size}
Now we compute $\mef$ for the DJIA stocks described above and
also for the S\&P~500 index where we use the approximately
16~year period from 2nd January~1992--15th February~2008 and
those 338 stocks out of the current 500 which were quoted in the
stock exchange during the whole period. After the correlation
matrix $\C$ is estimated, the effective portfolio size can be
calculated using $\C^{-1}$ and Eq.~\req{m_ef-MV-general} or
Eq.~\req{m_ef-Kelly-general}, it can be approximated using a
sector division and Eq.~\req{m_ef-sectors}, or it can be
approximated using $\avg{C}$ and Eq.~\req{m_ef-avgC}. We use the
division into nine sectors obtained from
\texttt{http://biz.yahoo.com/p} with the industry sectors: basic
materials, conglomerates, consumer goods, finance, healthcare,
industrial goods, services, technology, and utilities.
Using a different sector division (obtained \emph{e.g.} by
minimizing the ratio of average intra- and inter-sector
correlations) does not influence the results substantially. To
obtain the dependency of $\mef$ on the portfolio size $M$, we
select a random subset of $M$ stocks from the complete set and
compute $\mef$ for this subset; sensitivity to the subset
selection is eliminated by averaging over 5\,000 random draws.

\begin{figure}
\centering
\includegraphics[scale=0.28]{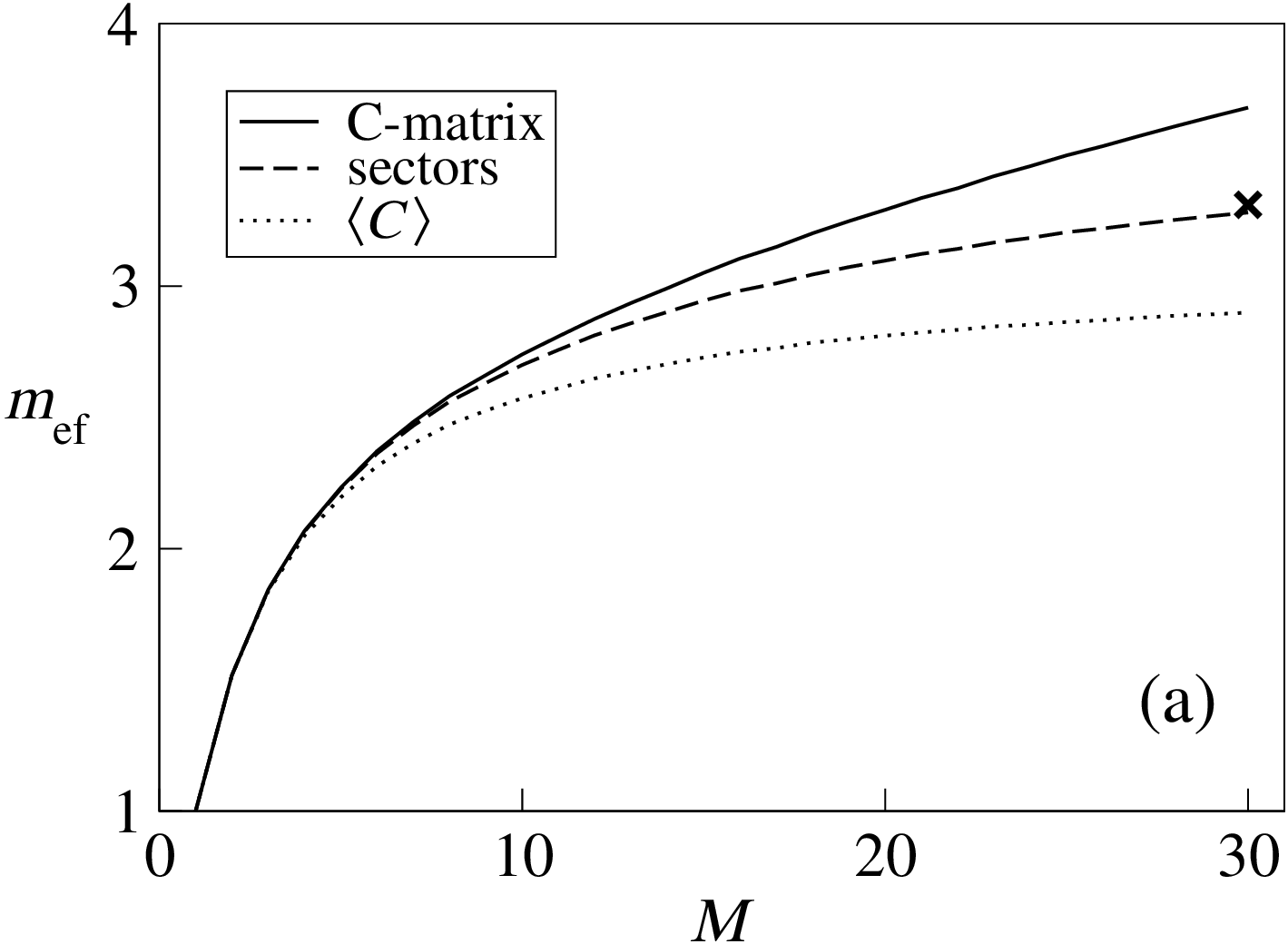}\quad
\includegraphics[scale=0.28]{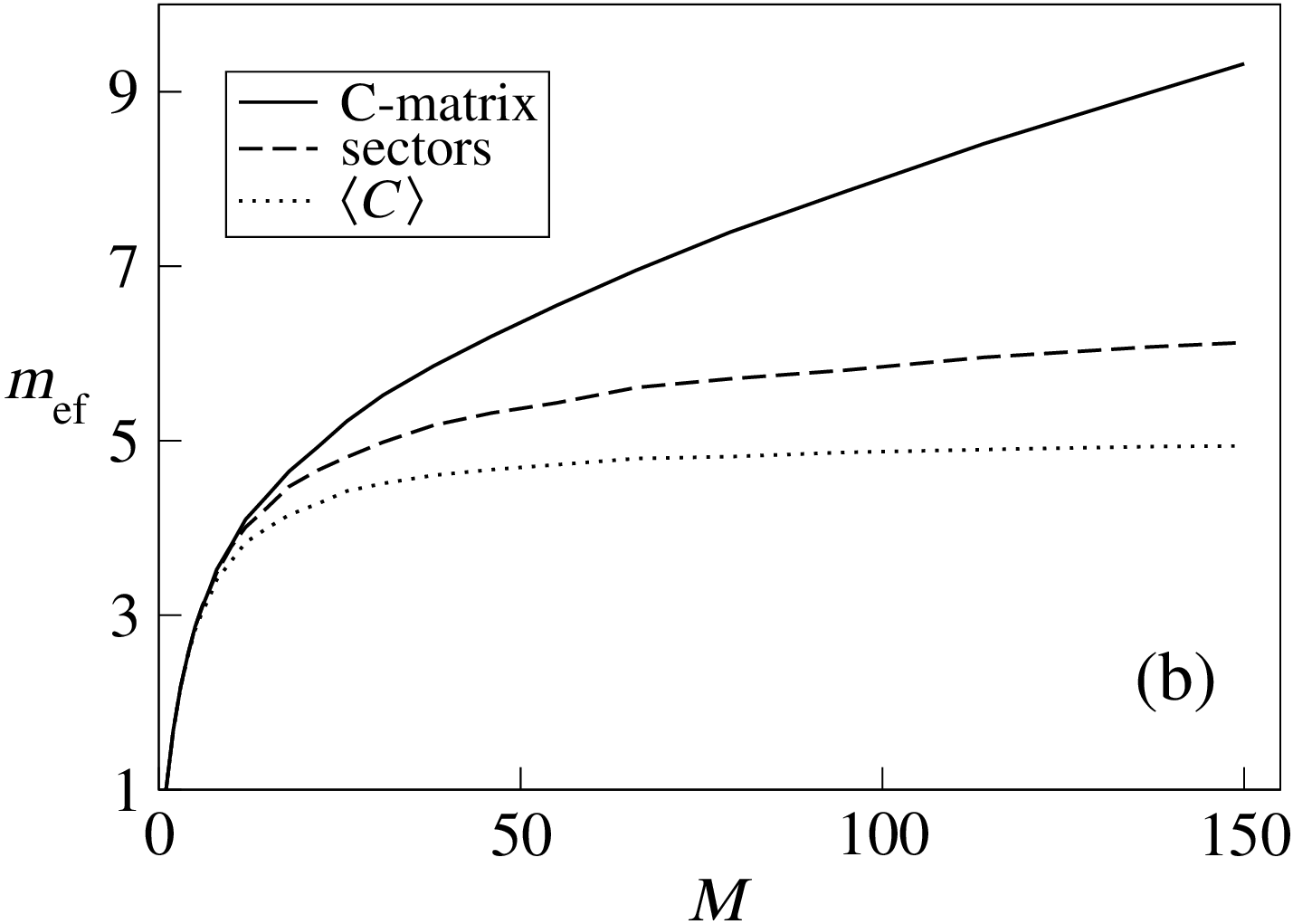}
\caption{The effective size $\mef$ for the DJIA stocks (a) and
the S\&P~500 stocks (b), computed directly from $\C$, from
sector division, and from $\avg{C}$. The estimate of the
effective size obtained by comparison of variances in
Sec.~\ref{sec-avgC} is shown as a thick cross. Individual
realizations fluctuate around the plotted averages
with the amplitude approximately one or less.}
\label{fig:m_ef-real}
\end{figure}

The results of the described analysis are shown in
Fig.~\ref{fig:m_ef-real}. For small portfolio sizes (less than
approximately ten stocks), the estimates of $\mef$ based on the
sector structure or on the average correlation perform well. For
a larger portfolio, the sector structure gives a better
description of $\mef$ than the average correlation.
Nevertheless, both estimates quickly saturate while $\mef$
obtained from the complete correlation matrix continues to grow
even for $M>100$. We see that the effect of heterogeneous
correlations increases with $M$ and even the sector structure is
insufficient to describe the system. The limited horizontal
scale in Fig.~\ref{fig:m_ef-real} is due to noisy estimates of
large correlation matrices from the data with a finite time
horizon $T$. As we are determining $M(M-1)/2$ correlations from
$MT$ prices, when $T$ is not very large compared to $M$, we face
an underdetermined system of equations~\citep{La99} (the problem
is also known as the dimensionality curse). With more frequent
financial data, larger portfolios would be easily accessible.

There is one particular point to be highlighted. According to
Eq.~\req{m_ef-even}, the estimate of $\mef$ by the average
correlation $\avg{C}$ is equal to the effective size $\mef'$ of
the portfolio containing all available assets with even weights.
Such investment was investigated \emph{e.g.} in~\citep{EG77}
where it was suggested that the benefits of diversification are
exhausted already with a portfolio of ten assets. This result
is in agreement with Fig.~\ref{fig:m_ef-real} where for both
sets of stocks, the effective portfolio size based on $\avg{C}$
saturates at $M\approx10$. However, since $\mef'$ is lower than
the effective sizes obtained directly from a sector division and
much lower than those obtained directly from $\C$, we can
conclude that with an evenly distributed investment one cannot
fully exploit the benefits of diversification.

\begin{figure}
\centering
\includegraphics[scale=0.28]{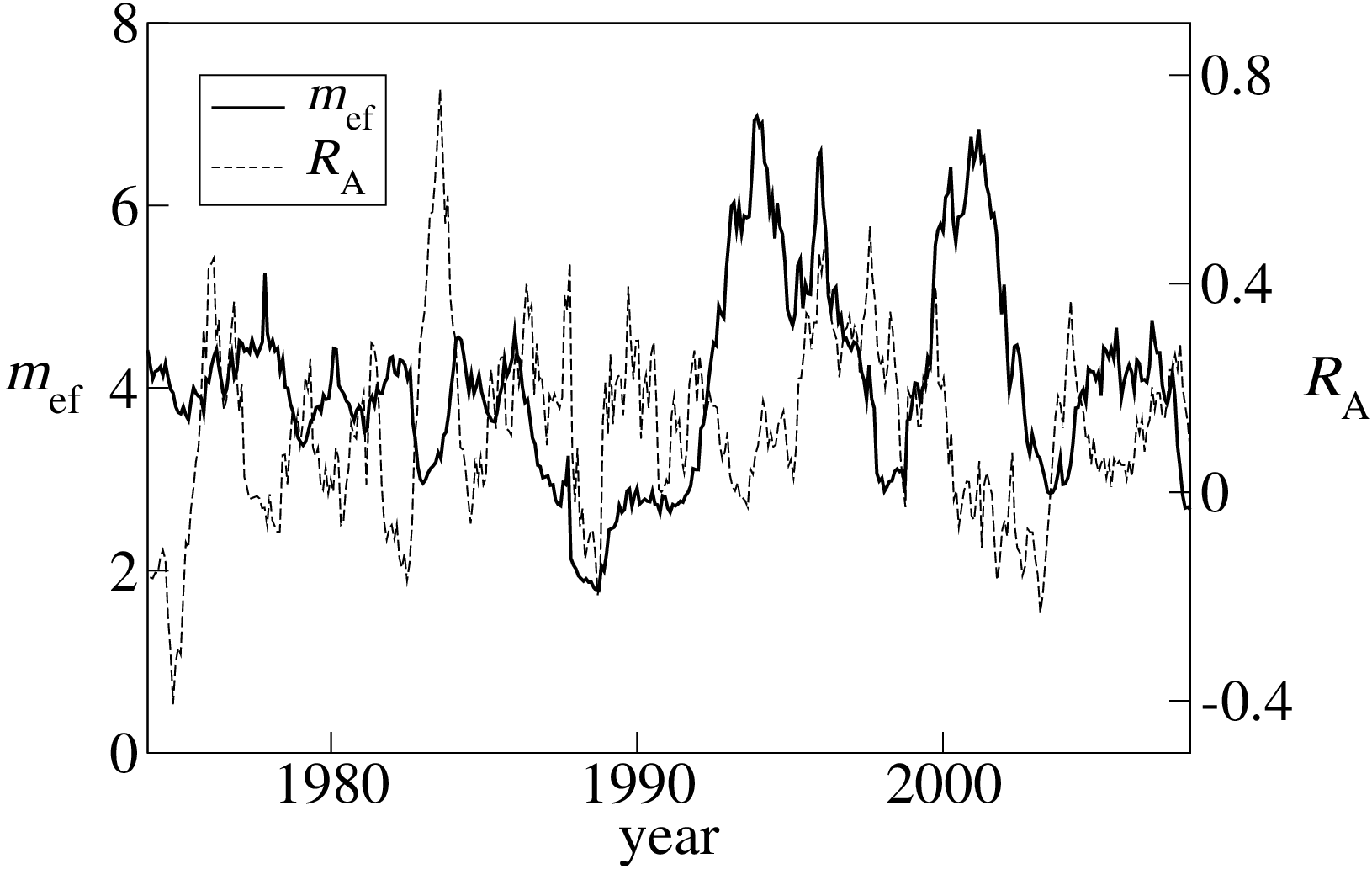}
\caption{Time evolution of the effective size $\mef$ and the
average yearly stock return $R_A$ for the DJIA stocks, based on
the sliding window of one year length. Note the three dips
corresponding to the black October of 1987, the emerging market
crisis of 1997, and the dot-com bubble bursting in 2001-2002.}
\label{fig:m_ef-evolution}
\end{figure}

Finally, in Fig.~\ref{fig:m_ef-evolution} we show the time
evolution of $\mef$ for twenty current stocks from the
DJIA.\footnote{The selected stocks are AA, BA, CAT, DD, DIS, GE,
GM, HON, HPQ, IBM, JNJ, KO, MCD, MMM, MO, MRK, PG, UTX, WMT, and
XOM.}
We use the daily data from the period Jan 1973--Apr 2008 and the
sliding window of one year length to obtain estimates of the
correlation matrix and compute $\mef$ from $\C^{-1}$. In the
same figure we show also the average yearly return of the
selected stocks (estimated on the one-year basis). As we see,
$\mef$ varies with a large amplitude, being as low as two at the
end of 80's (corresponding to the October crash of 1987) and
peaking at almost seven during the year 1994.

\section{Conclusion}
In today's globalized world where a single event can have
world-wide implications, even internationalized portfolios are
not immune to asset correlations. As a result, measurement and
effects of correlations remain as a prominent challenge of
portfolio theory. In this work we focused on the influence of
correlations on the performance of optimal portfolios. On
a~simple artificial example we showed that underestimation of
asset correlations can lead to a significant reduction of
profitability (see Fig.~\ref{fig:G}).

Even when correlations are correctly estimated, they harm the
investment performance by reducing the true degree of
diversification. To measure their influence we introduced a new
quantity: the effective portfolio size $\mef$. We derived simple
analytical expressions which allow us to easily calculate $\mef$
both for the Mean--Variance portfolio and the Kelly portfolio.
We showed that with increasing number of stocks, the
diversification measure increases only slowly and in some cases
it even saturates without any further net diversification
effect. In agreement with previous studies, evenly distributed
investment turns out to be a rather ineffective way of
diversification as it results in relatively small values of
$\mef$. In addition, the time dependence of $\mef$ obtained
from prices of the DJIA stocks shows strong its heavy variations
and also minima corresponding to the crashes that occurred in the
investigated period.

Although illustrated only on the Mean--Variance portfolio and
the Kelly portfolio, the effective portfolio size is a~general
concept which can be used also in other optimization methods. It
reduces the complex structure of the full correlation matrix to
a~single number with a simple interpretation and thus it allows
us to appreciate how much do correlations harm our investment.
In future research, numerical results for $\mef$ should be
refined using high frequency financial data which would allow
for shorter sliding window lengths and fewer averaging
artifacts. Since correlation estimates are noisy even with
extensive data available, it would be interesting to see how the
concept of the effective size applies to a noisy correlation
matrix. Eventual direct applications of the proposed concept to
portfolio management remain as a future challenge.

\section*{Acknowledgments}
We acknowledge useful discussions with Damien Challet and early
mathematical insights of Jozef Mi\v skuf. This work was
supported by the Swiss National Science Foundation (Project
205120-113842) and in part by SBF Switzerland (project
No.~C05.0148 Physics of Risk) and by the International
Scientific Cooperation and Communication Project of Sichuan
Province in China (Grant No.~2008HH0014). C.~H.~Yeung
acknowledges the hospitality of the University of Fribourg, the
ORA of HKUST, and the support by the Research Grant Council of
Hong Kong (Grant Numbers HKUST603606 and HKUST603607).

\end{document}